# A machine learning approach to investigate regulatory control circuits in bacterial metabolic pathways.

Francesco Bardozzo[(1)], Pietro Liò [(2)], Roberto Tagliaferri [(1)]

(1)NeuRoNe Lab, DISA-MIS, University of Salerno,
Via Giovanni Paolo II, 132, Fisciano, Italy , robtag@unisa.it

(2) Computer Laboratory, University of Cambridge,
15 JJ Thomson Ave, CB3 0FD, Cambridge, United Kingdom, pl219@cam.ac.uk

**Abstract.** In this work a machine learning approach for identifying the multi-omics metabolic regulatory control circuits inside the pathways is described. Therefore, the identification of bacterial metabolic pathways that are more regulated than others in term of their multi-omics follows from the analysis of these circuits . This is a consequence of the alternation of the omic values of codon usage and protein abundance along the circuits. In this work, the E.Coli's Glycolysis and its multi-omic circuit features are shown as an example.

## 1 Background

In the bacterial metabolic pathways, it is possible to identify different small circuits that lead from an intermediate compound to another. Each bacterial pathway could be considered as a highly specific directed graph that presents more than one multi-omic circuit (MOC). In standard conditions is possible to identify which pathways are more regulatory than others in terms of their alternating multi-omic contribution. The MOCs that belong to specific pathway could be discovered through the flux-balance analysis [1]. On the other hand, in this work, we propose a machine learning approach to study the MOCs. This method takes into account multi-omic values and looks for the information derived from alternate sequences of multi-omic values. The proteins are ordered in a sequence with respect to their position on the circuit. Omic values of protein abundance and codon usage are associated with these proteins. An upper bound corresponding to the ideal sequence of alternated omic values is given. Then, the alternated sequence of omic values is compared with the ideal alternation. It is possible that the presence of alternated omic values reflects the metabolic regulatory control behaviour of the specific circuit inside the metabolic pathways. The more the alternated values in the sequence are different from the upper bound, the less the circuit is regulated and vice versa [2]. Another important consequence strictly related to the identification of the MOCs is concerning the importance of identifying the intermediate compound in the circuit output. For example, the intermediate compound at the end of a circuit could be considered as a result of a regulatory control path and could have for this reason a strategical importance in the design of a metabolic network. In this setting, this work may be useful in the metabolic networks reconstruction based on metabolic functional data [3, 4].

## 2 Materials and Methods

Our analyses are based on the Glycolysis of Escherichia coli K-12 MG1655. The codon adaptation index (CAI) is a codon usage index computed as described by Sharp and Li [5]. The second omic value considered is the protein abundance (PA) in standard conditions and is extracted by the PaxDb database [6]. PA and CAI are two omics that could be considered correlated. In fact, in an integrated analysis of multi-omic MOCs, the CAI stabilises the measures variability of PA, that in turn presents a correlation with the mRNA transcript [9]. The information about genes, proteins and metabolic pathways is extracted from NCBI Gene Bank [8] and KEGG [7].

### 2.1 Machine Learning Approach

An MOC presents a starting point, that we could call *starting protein* (SP) and an end point, that we could call *ending protein* (EP). Between this two points a different number of proteins with their own PA and CAI could be present.

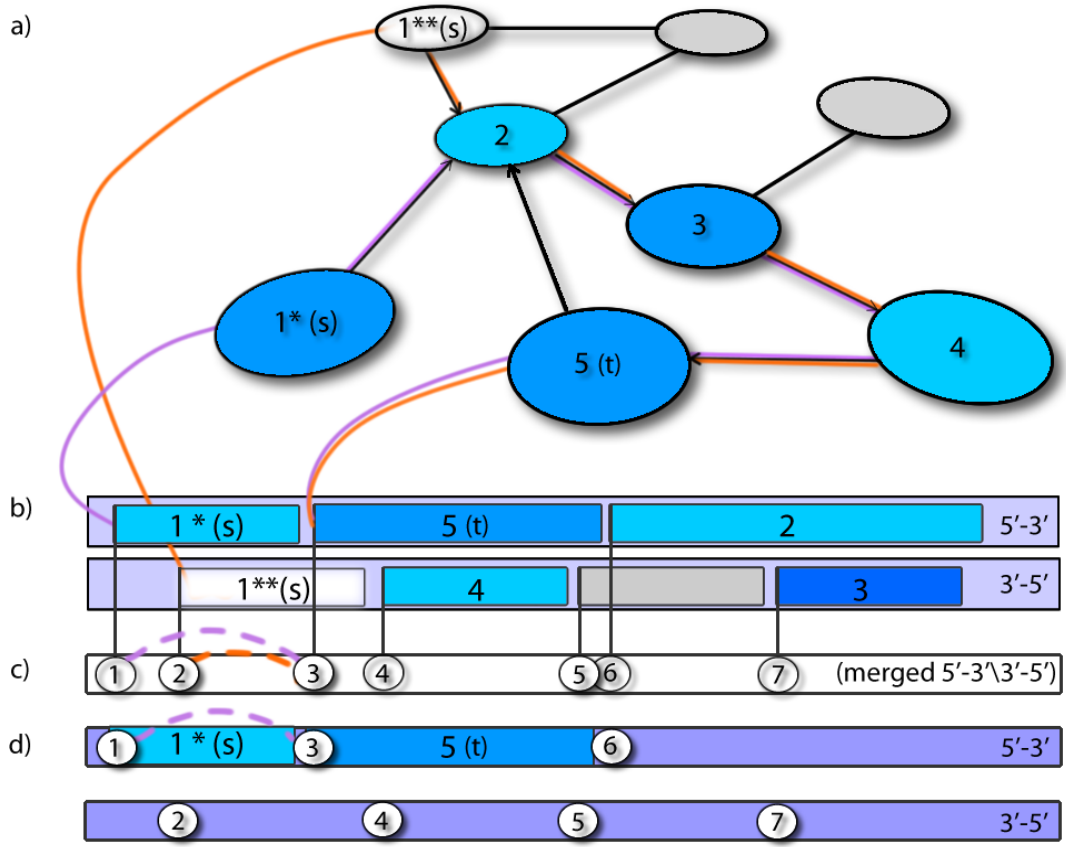

Figure 1: (a) In this metabolic pathway 2 MOCs of length 5 are found. (b) The position of the genes on the double strand and the proteins codified at the network level are shown. (c) The two MOCs are selected with 2 different criteria. The former is based on the selection of SP and EP merging the double strand and considering the positions of the genes as merged. The latter selects the SP and EP from the strand (5'-3') or from the strand (3'-4'). In this last is impossible to have a circuit that has an SP from the 5'-3' strand and an EP from the 3'-5' strand.

In Figure 1 (a) two circuits individuated from two different SPs (s) (1** and 1*) and that lead to a single EP (5 (t)) are shown. We can see that the EP is chosen as the end of the circuit because there is a directed path from the SPs to this protein. Obviously, we are maintaining the order of the genes (related to these proteins) on the double strand while we identify the SP and EP . For example, in Figure 1 (b) both the 1* and the 1** genes are positioned before the gene labelled as 5 (t). The found paths (violet and orange) are the shortest paths of length 5 but, in this case, there is a relevant difference between the position of the SP (s). In fact, the first SP is positioned in the same direction (5'-3') of the EP. Instead, the second one is positioned in the direction 3'-5' with respect to the (5'-3') EP. This exchange of direction between the genes may be also relate to genes that are located in between the circuit and not only at the extremes. Therefore, it is of fundamental interest to find if the genes that are on the same DNA strand produce proteins that are belonging to the same MOC. Moreover, this is related to the way the MOCs are formed: if their genes are in the same 5'-3' or 3-5' single strand or alternating in the double strand. This information could be helpful to the purpose of understanding the metabolic regulatory control. As described in Figure 1 (c) we can consider a sort of merged double strand into a single one. As a consequence, all the couples of SP and EP are organised in a single sequence and this merged strand contains 2 MOCs (orange and violet). On the other hand, if we consider each single strand separately we can have only a MOC, the violet one (Figure 1 (d)). In particular, the typical bacterial polycistronic organisation suggests that each polycistronic mRNA carries the information of more than one gene [10]. In effect, the genes are organised in operons and for this reason are located on the same single DNA strand. Therefore, these operons codify enzymes with related functions and frequently are involved in the same metabolic pathway. This type of characteristic reflects the functional necessity of rapidly responding to the external environment, activating a particular metabolic pathway. This might suggest that, when we think about MOCs, the model that considers the SP and the EP that comes from the same strand (Figure 1 (d)), could be better than the model that consider these points in a merged DNA strand (Figure 1 (c)). The objective of this work is to propose a method to find MOCs and to show they are relevant in a metabolic pathway. An exhaustive search for selecting all the couples of SP-EP is applied. Then the models composed by a merged single strand and considering the strands separately are built. As illustrated in [11] it is possible to find a function $\Psi$ to compute the number of classes for the normalised multi omic values of the MOCs. In this way, as illustrated in Figure 2 we can transform the multi omic values into a vector of integers $v$.

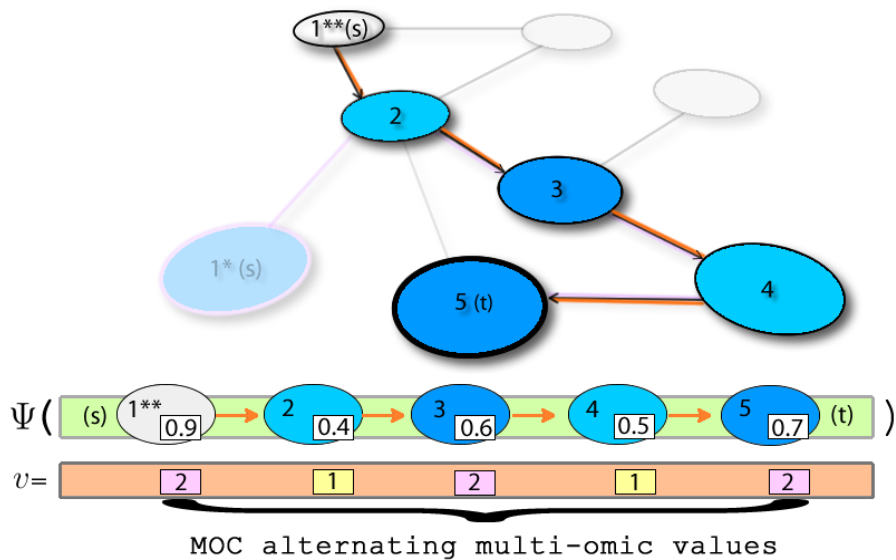

Figure 2: The multi-omic values of the orange MOC are transformed in integers by the function $\Psi$. This circuit presents alternating multi-omic values ($v$).

For each $v$ we can obtain a score of the relative distances between the omics in their incremental position, as described in [12]. Then, it is possible to obtain a similarity measure $\sigma$ between the scored MOC's $v$ and the ideal sequence. Figure 3 plots the similarity measures against the circuit lengths, from zero to one corresponding to increasing size dots. Moreover, in Figure 3 the proteins produced by a single DNA strand are coloured in yellow, those produced by double strands are in violet.

## 3 Results

We have extracted all the possible MOCs from the Glycolysis. $\Psi$ returns the number of classes equal to 7. The number of proteins of this metabolic pathway is of 40. We computed for all the MOC's sequences the score for the multi-omic contributions of the CAI and PA. All the values before being transformed by $\Psi$ are normalised with a standard normalisation, summed in and averaged. The metabolic pathway of Glycolysis presents 134 potential MOCs, in the direction from the SP to the EP.

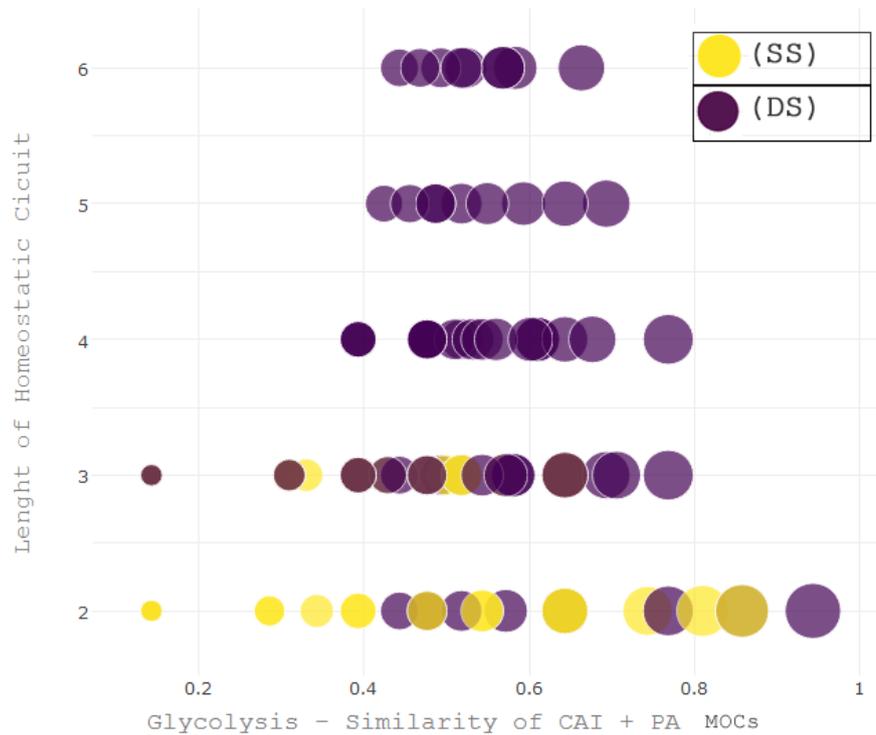

Figure 3: The plot shows the circuits extracted from the Glycolysis metabolic pathway. The similarity $\sigma$ is displayed on the axis x, while the circuit length is on the axis y. Furthermore, this plot underlines the relationship between the circuit length and its similarity $\sigma$ to the ideal alternate sequence of multi-omic values. The similarity increases with the size of the dots. The proteins produced by a single DNA strand (SS) are coloured in yellow, those produced by double strands (DS) are in violet

In particular, in Figure 3 the MOC of Glycolysis presents a length varying from 2 to 6 proteins. The number of single strand MOCs on the Glycolysis are 47 over 134 MOCs in total. The yellow dots in Figure 3 are only on MOCs of length 2 and 3. In the Glycolysis, there are only 4 operons, and only two of them form two MOCs that cover the proteins aceE, aceF, lpd and ascF, ascB, with a $\sigma \leq 0.25$ In this example there is not any operon that in the metabolic network represents a MOC with a high $\sigma$.

## 4 Conclusion

We presented a machine learning approach for the individuation of metabolic regulatory control circuits inside the bacterial metabolic pathways. The MOCs were investigated in relation to multi omics data. We have shown that the proteins related to the operons have not a key role when their proteins are present in the MOCs. Moreover in this pathway, a different distribution with respect to the length of the MOCs between the single strand and double strand models is showed. Obviously, is necessary to extend the analyses to the whole genome and to the study the variations caused by external factors.